\documentclass[fleqn,usenatbib,useAMS]{mnras}

\usepackage{newtxtext,newtxmath}

\usepackage[T1]{fontenc}
\usepackage{ae,aecompl}
\usepackage{xcolor}

\usepackage{graphicx}   
\usepackage{amsmath}    
\usepackage{relsize}    
\usepackage{subcaption}

\renewcommand{\vec}[1]{\boldsymbol{#1}}
\newlength{\VSpaceBeforeTabBib}
\newlength{\VSpaceBeforeTabFoot}
\usepackage{xcolor}
\usepackage{xifthen}
\usepackage[normalem]{ulem}
\newcommand{\stkout}[1]{\ifmmode\text{\sout{\ensuremath{#1}}}\else\sout{#1}\fi}
\newcommand{\edited}[2]{\ifthenelse{\isempty{#1}}{\textcolor{red}{#2}}{\ifthenelse{\isempty{#2}}{\textcolor{gray}{\stkout{#1}}}{\textcolor{gray}{\stkout{#1}} \textcolor{red}{#2}}}}

\title{Centrifugal Barrier and Super-Keplerian Rotation in Protostellar Disk Formation}
\author[Authors]{
Dylan C. Jones,$^{1}$\thanks{dylanjonesc@gmail.com}
Ka Ho Lam,$^{1}$
Zhi-Yun Li,$^{1}$
Yisheng Tu$^{1}$
\\
$^{1}$Department of Astronomy, University of Virginia, Charlottesville, VA 22904, USA\\
}

\date{Accepted XXX. Received YYY; in original form ZZZ}

\pubyear{2019}

\begin{document}
\label{firstpage}
\pagerange{\pageref{firstpage}--\pageref{lastpage}}
\maketitle

\begin{abstract}
With the advent of ALMA, it is now possible to observationally constrain how disks form around deeply embedded protostars. In particular, the recent ALMA C$_3$H$_2$ line observations of the nearby protostar L1527 have been interpreted as evidence for the so-called ``centrifugal barrier,'' where the protostellar envelope infall is gradually decelerated to a stop by the centrifugal force in a region of super-Keplerian rotation. To test the concept of centrifugal barrier, which was originally based on angular momentum conserving-collapse of a rotating test particle around a fixed point mass, we carry out simple axisymmetric hydrodynamic simulations of protostellar disk formation including a minimum set of ingredients: self-gravity, rotation, and a prescribed viscosity that enables the disk to accrete. We find that a super-Keplerian region can indeed exist when the viscosity is relatively large but, unlike the classic picture of centrifugal barrier, the infalling envelope material is not decelerated solely by the centrifugal force. The region has more specific angular momentum than its surrounding envelope material, which points to an origin in outward angular momentum transport in the disk (subject to the constraint of disk expansion by the infalling envelope), rather than the spin-up of the envelope material envisioned in the classic picture as it falls closer to the center in order to conserve angular momentum. For smaller viscosities, the super-Keplerian rotation is weaker or non-existing. We conclude that, despite the existence of super-Keplerian rotation in some parameter regime, the classic picture of centrifugal barrier is not supported by our simulations. 
\end{abstract}




\section{Introduction} \label{sec:intro}






Disks play a central role in the formation of both stars and planets. How they form and evolve remain hotly debated. Part of the difficulty is that disk formation is strongly affected by magnetic fields, which are known to permeate molecular clouds and their star-forming cores but difficult to quantify \citep{2019FrASS...6...15P,2019FrASS...6....3H}. Theoretical calculations have shown that disks formed in strongly magnetized molecular cloud cores depend on a number of factors, including non-ideal MHD effects, turbulence, and rotation-magnetic field misalignment \citep[for reviews, see, e.g.,][]{2014prpl.conf..173L,2020SSRv..216...43Z}.
However, it has been challenging to  observationally constrain these calculations, because deeply embedded protostellar disks are still in active formation and their connection to the protostellar envelopes are difficult to characterize through observations. 


The observational situation has improved drastically with the advent of ALMA. In particular, \citet{2014Natur.507...78S} traced the gas kinematics in the transition zone between the infall envelope and the rotationally supported disk in the well-studied L1527 protostellar system. They found that the position-velocity (PV) diagram of cyclic-C$_3$H$_2$ along the equator of the system can be fitted remarkably well by a simple analytic model first introduced by  \citet{ulrich1976infall} and \citet{cassen1981formation}. This model describes the motion of a test particle in the gravity field of a central stellar object of a fixed mass $M_*$ under the constraint of energy and angular momentum conservation. It is characterized by two special locations: the so-called ``centrifugal radius'' ($r_\mathrm{CR}$) and ``centrifugal barrier'' ($r_\mathrm{CB}$). The former is reached when the rotational speed $v_\phi$ of the test particle reaches the (local) Keplerian speed $v_\mathrm{K}= (G M_* / r)^{1/2}$, where $r$ is the distance of the particle from the star. Outside the centrifugal radius $r_\mathrm{CR}$, the gravity dominates the centrifugal force, leading to a faster and faster collapse towards the central object and a maximum infall speed at the centrifugal radius. Inside $r_\mathrm{CR}$, the rotation becomes super-Keplerian, with the centrifugal force dominating the gravity, which leads to a deceleration of the infall motion. The infall comes to a complete stop at the centrifugal barrier $r_\mathrm{CB}$ (which is located at a radius half of that of the centrifugal radius, i.e., $r_\mathrm{CB}=r_\mathrm{CR}/2$), where the rotation speed is $\sqrt{2}$ times the local Keplerian speed. 

The apparent agreement between the analytic model and the gas kinematic data was taken as evidence for the existence of a centrifugal barrier \citep{2014Natur.507...78S,sakai2017vertical}, where the envelope infall is completely stopped by the centrifugal force from a super-Keplerian rotation. If true, it would be evidence that the infall in the envelope comes to a stop gradually over an extended radial zone (between $r_\mathrm{CR}$ and $r_\mathrm{CB}$) rather than abruptly at an accretion shock where the rapidly infalling envelope material slams into the rotationally support disk, which would have far reaching consequences on the structure and dynamics of the envelope-to-disk transition zone, particularly its temperature structure and chemistry. 

However, the concept of centrifugal barrier is based on the assumption that the envelope material moves as (non-interacting) test particles. The lack of gas pressure under this assumption means that the infall of the material on the equatorial plane can be stopped only by super-Keplerian rotation; those from above and below the equatorial plane are presumed to terminate on the equatorial plane (see Fig.~2 of \citealp{cassen1981formation} for an illustration) before they are stopped by the super-Keplerian rotation, so that the concept of centrifugal barrier is not directly applicable for them. Even for the material moving on the equatorial plane, the centrifugal barrier is not an equilibrium location, since the inward gravitational pull is too weak to balance the outward centrifugal force associated with the super-Keplerian rotation. {\color{blue} Indeed, it is maximally out of equilibrium.} Under the idealization on which the centrifugal barrier is based, the (non-interacting) test particles would move outward after reaching the centrifugal barrier. The expanding material is expected to collide with the material that continues to fall inward, leading to (gas pressure-mediated) modifications that cannot be captured by the test-particle approach. Whether the classic picture of centrifugal barrier can survive in a more realistic fluid approach is unclear. It is the main topic of our investigation. 

The rest of the paper is organized as follows. We start with problem setup in \S~\ref{sec:setup}. It is followed by numerical results in \S~\ref{sec:results}, where we find that super-Keplerian rotation indeed exists in the outer part of the disk that separates the Keplerian part of the disk from the infall envelope when the viscosity is relatively high, although its physical origin is distinct from the centrifugal barrier, and that it weakens with decreasing viscosity. The classic picture of centrifugal barrier is not supported by our simulations. We discuss the results and conclude in \S~\ref{sec:discussion}.

\section{Problem Setup}
\label{sec:setup}

\subsection{Governing Equations}

As mentioned in \S~\ref{sec:intro}, protostellar disk formation in dense cores of molecular clouds is a complex process involving turbulence and magnetic fields (and associated non-ideal MHD effects). To check whether the centrifugal barrier plays a role in disk formation in the fluid approach, we have decided to limit our investigation to the simplest case of the collapse of a rotating core without any magnetic field or turbulence; these neglected effects will be discussed in \S~\ref{sec:discussion}. The core collapse and disk formation is governed by the continuity equation 
\begin{equation}
  \frac{\partial \rho}{\partial t} + \nabla \cdot \left( \rho \vec{v} \right) = 0,
\end{equation}
and the momentum equation
\begin{equation}
  \label{eq:momentum}
  \rho \frac{\partial \vec{v}}{\partial t} + \rho \left( \vec{v} \cdot \nabla \right) \vec{v} = -\nabla P - \rho \nabla \Phi_\mathrm{g} + \nabla\cdot \vec{\Pi},
\end{equation}
where the gravitational potential is evaluated using the Poisson equation
\begin{equation}
  \nabla^2 \Phi_\mathrm{g} = 4 \pi G \rho.
\end{equation}
and the tensor $\vec{\Pi}$ is determined by the coefficient of shear viscosity $\rho \nu$ \citep[see, e.g., equation~7 of][]{2020ApJS..249....4S}. We include a non-zero shear viscosity so that a disk rather than a ring is formed \citep[e.g.][]{kuiper2010circumventing} and the formed disk can evolve even in the absence of a magnetic field. Following \citet{kuiper2010circumventing},  we will adopt the standard $\beta-$prescription for the kinematic viscosity $\nu$ (not to be confused with the velocity $\vec{v}$):
\begin{equation}
    \nu = \beta \Omega_\mathrm{K}(r) R^2, 
    \label{eq:viscosity}
\end{equation}
where $R$ is the cylindrical radius and $\Omega_\mathrm{K}(r)=\sqrt{GM(r)/r^3}$, with $M(r)$ denoting the mass enclosed within a sphere of radius $r$. For a rotationally supported disk, the quantity $\Omega_\mathrm{K}$ would be the orbital angular velocity and the dimensionless parameter $\beta$ would be related to the standard $\alpha-$viscosity parameter by $\beta=\alpha (H/R)^2$ where $H$ is the disk scale height. The viscosity is applied throughout the simulation domain. It has relatively little effect at large radii well beyond the disk formation region.

\subsection{Numerical Method}

The governing equations are solved with the \texttt{Athena++} code \citep{2020ApJS..249....4S} under the assumption of (2D) axisymmetry around the rotation axis in a spherical polar coordinate system $(r,\theta,\phi)$. We adopt a computation grid with logarithmic spacing in the radial direction and constant spacing in the theta direction. Logarithmic spacing has the advantage of providing higher resolution at smaller  radii relevant to disk formation, at the expense of producing larger cells at larger radii, which are not as critical for our analysis. 

%
%

There are 4 boundaries where boundary conditions need to be imposed in the simulation. The inner and outer radial boundaries are treated the same, with a semi-outflow boundary condition where material is allowed to exit the computational domain but not enter back in. The standard reflective boundary condition is imposed at the poles.  

\section{Results}
\label{sec:results}

\subsection{Reference Model} 
\label{sec:reference}




We will start our discussion with a reference model that shows prominent super-Keplerian rotation that allows us to analyze its origin. This is followed by a set of simulations that illustrate its dependence on model parameters, especially the viscosity. 

Our reference model has an initial uniform density profile with an enclosed mass of $4.2~\mathrm{M_\odot}$, an isothermal sound speed of $c_s = 0.2~\mathrm{km\,s^{-1}}$, and a $\beta-$viscosity parameter of $0.077$. The inner and outer radii of this model are, respectively, 2~AU and 10,000~AU. The polar angle $\theta$ ranges from $0$ to $\pi$. An initial solid body rotation of $\Omega = 5.945 \times 10^{-14}~\mathrm{rad\,s^{-1}}$ was used, corresponding to a dimensionless ratio of rotational and gravitational energies of $\beta_\mathrm{rot} \approx 0.7\%$. As often done in 
disk formation simulations \citep[e.g.][]{tomida2017granddesign}, we change the isothermal equation of state to a stiffer, adiabatic, equation of state (with an adiabatic index of 5/3) around a critical mass density of $\rho_\mathrm{c}=10^{-13}~\mathrm{g\,cm^{-1}}$),  namely, 
\begin{equation}
P = \rho c_s^2 \left[ 1 + \left( \frac{\rho}{\rho_\mathrm{c}} \right)^{2/3}  \right] \label{eq:eos}
\end{equation}
where $c_s$ is the isothermal sound speed. 

The reference simulation was carried out on a $512\times256$ ($r$-$\theta$) grid, which ensures that the cells are roughly square-shaped. We have also experimented with lower and higher resolution simulations, with similar results. 

\begin{figure*}
    \centering
    \includegraphics[{trim=2cm 2cm 2cm 2cm},clip,width=\textwidth]{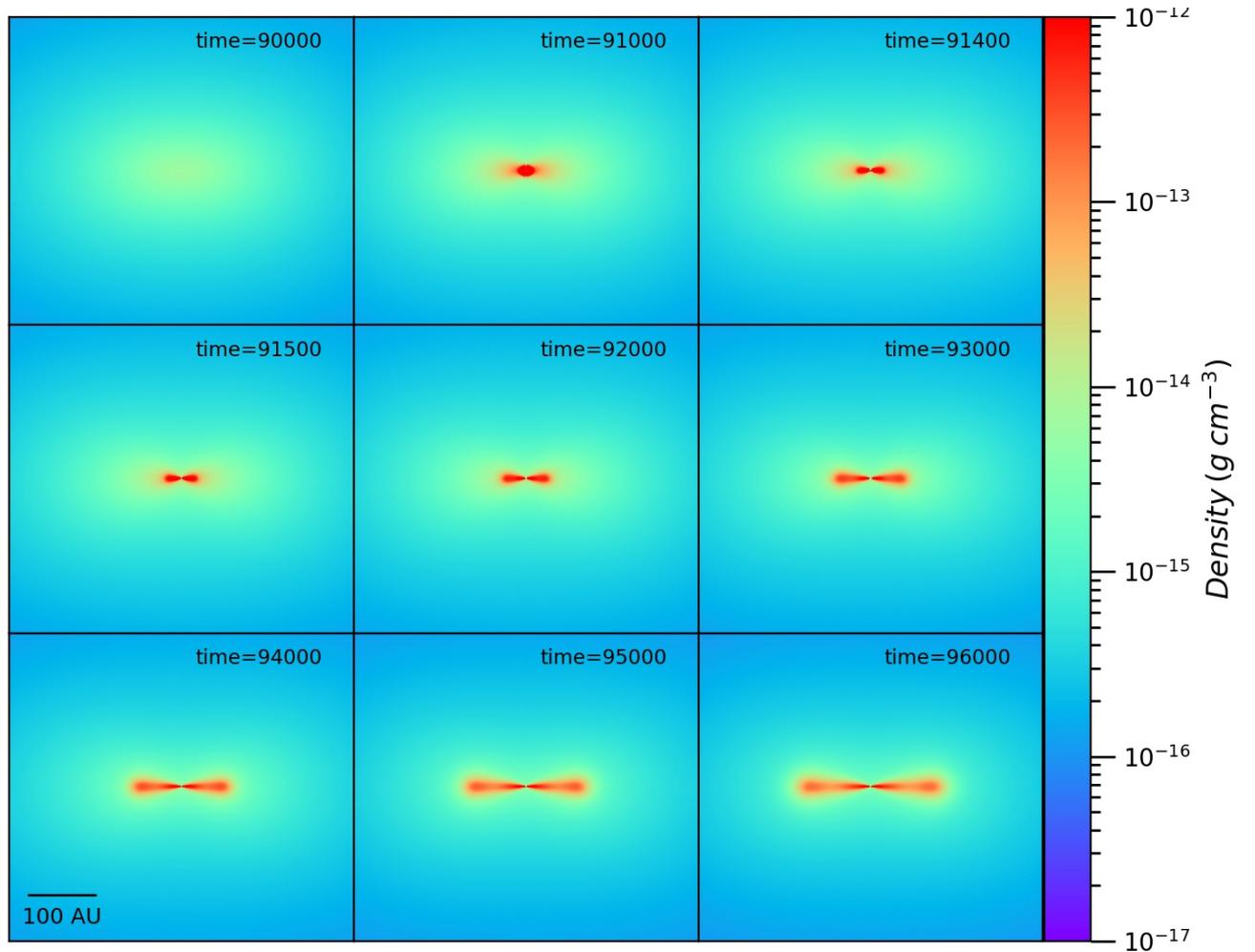}
    \caption{Evolution of the density distribution of our reference model showing the formation and growth of a disk. The units for the time and density are, respectively, years and $\mathrm{g\,cm^{-3}}$.}
    \label{fig:fig1}
\end{figure*}

\begin{figure}
    \centering
    \includegraphics[clip,width=\columnwidth]{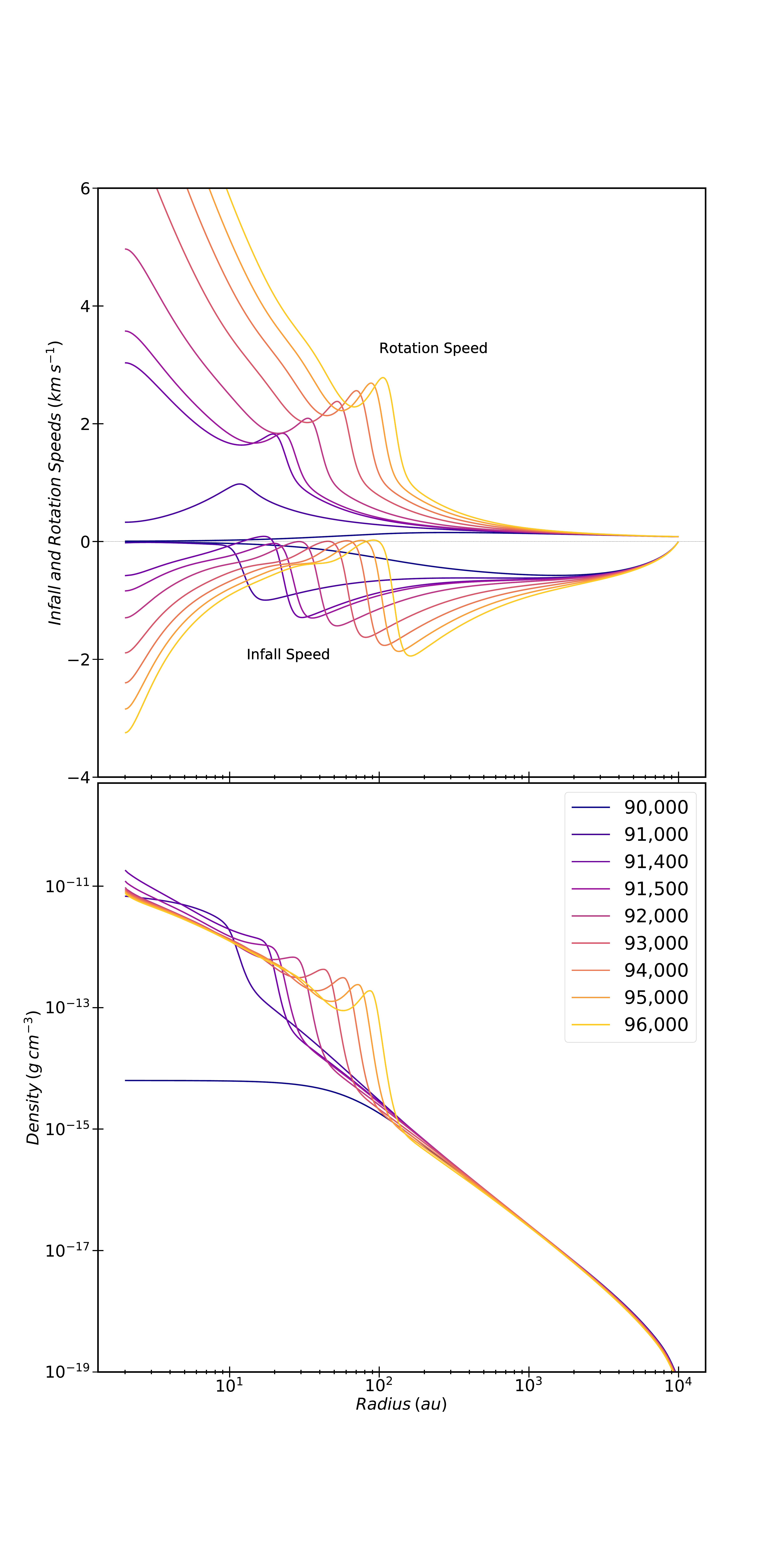}
    \caption{Evolution of the infall and rotation speeds and density on the equatorial plane for the frames shown in Fig.~\ref{fig:fig1}.
    }
    \label{fig:fig2}
\end{figure}

The dynamic evolution of the system can be divided into two phases, as illustrated in Fig.~\ref{fig:fig1}. In the first, pre-stellar, phase, the slowly rotating dense core material collapses towards the center, with a flat distribution of density that increases with time. By $t=91,000$~years after the start of the collapse, the density in the central region becomes high enough that a thermally-supported first core with a radius of 20~AU forms (see the top-middle panel). Following the first core, a rotationally supported disk forms which grows in radius with time for the remainder of the simulation. At the last frame shown ($t=96,000$ years), we form a disk with the radius on the order of 120~AU. This progression within our reference model can also be seen within the time sequence of the spatial distributions of the density, infall and rotation speeds on the equatorial plane in Fig.~\ref{fig:fig2}. It shows clearly the outward propagation of the transition region from the infall-dominated lower-density envelope to the rotation-dominated denser disk.  

\begin{figure*}
    \centering
    \includegraphics[width=.98\linewidth]{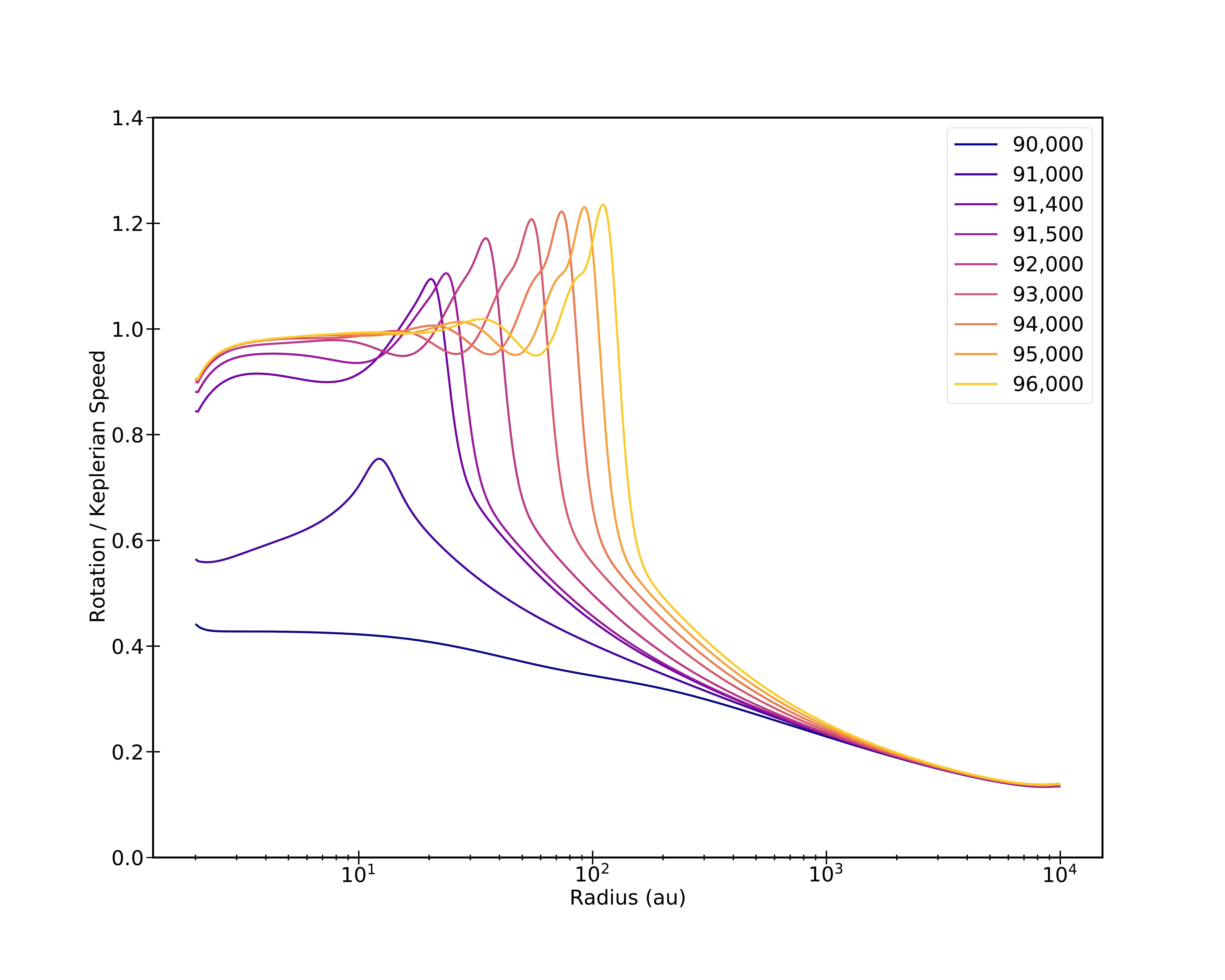}
    \caption{Ratio of the rotational speed to the Keplerian speed at the representative times shown in Figs.~\ref{fig:fig1} and \ref{fig:fig2}, showing the existence of super-Keplerian rotation.}
    \label{fig:vphi_ratio}
\end{figure*}

To address the question of whether a centrifugal barrier exists in our reference model or not, we plot in Fig.~\ref{fig:vphi_ratio} the ratios of the rotational speed $v_\phi$ and the local Keplerian speed $v_\mathrm{K}$ \footnote{The local Keplerian speed is defined as the rotation speed that yields a centrifugal force that balances the local gravitational force exactly.} on the equatorial plane at different times, which allows us to search for super-Keplerian rotation that is a key ingredient of the centrifugal barrier. 
It is clear that during the pre-stellar collapse phase the rotation remains sub-Keplerian, as illustrated by the curve for $t=90,000$~years. Inside the thermally-supported first core, the rotation remains sub-Keplerian, although it is fast enough to flatten the first core significantly (see Fig.~\ref{fig:fig1}, $t=91,000$~years). Before the thermally supported sub-Keplerian first core evolves into a rotationally supported Keplerian disk, super-Keplerian rotation develops right outside the first core (see the curve for $t=91,400$~years). It persists as the first core spins up to the Keplerian speed and evolves into the rotationally supported disk (see the curve for $t=91,500$~years). 
As the disk grows in time, it continues to be surrounded by a super-Keplerian region that expands with the disk. 
Once a well defined disk has formed at 100~AU, the location of maximum super-Keplerian rotation (i.e., the peak of the curve of rotation speed normalized by the local Keplerian speed in Fig.~\ref{fig:vphi_ratio} or super-Keplerian peak hereafter) is expanding outwards at a velocity on an order of $4 \times 10^4~\mathrm{cm\,s^{-1}}$, which is faster than the speed with which the material at the super-Keplerian peak is moving (inwards), which is of order $2 \times 10^4~\mathrm{cm\,s^{-1}}$. 
This difference between the expansion velocity and the local infall velocity is evidence that the super-Keplerian feature acts as a relatively fast outward-propagating wave against a slowly inward-drifting material.

The question that next naturally arises is: is this super-Keplerian region the centrifugal barrier that was introduced by \citet{ulrich1976infall} and \citet{cassen1981formation} and described in the introduction section? 

\begin{figure*}
    \centering
    \begin{subfigure}[b]{.49\textwidth}
        \centering
        \includegraphics[width=\linewidth]{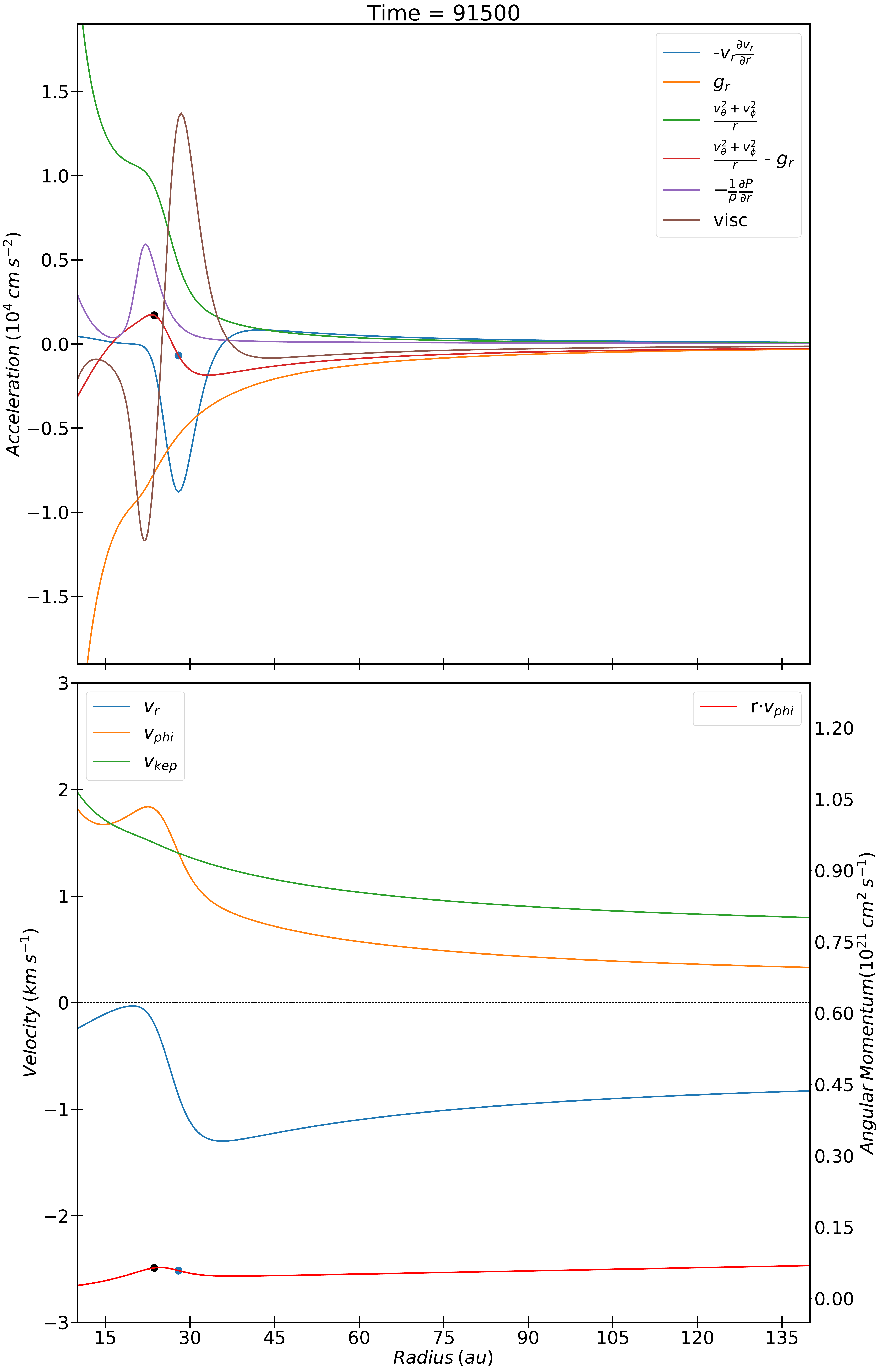}
        \label{fig:acag-first}
    \end{subfigure}
    \hfill
    \begin{subfigure}[b]{.49\textwidth}
        \centering
        \includegraphics[width=\linewidth]{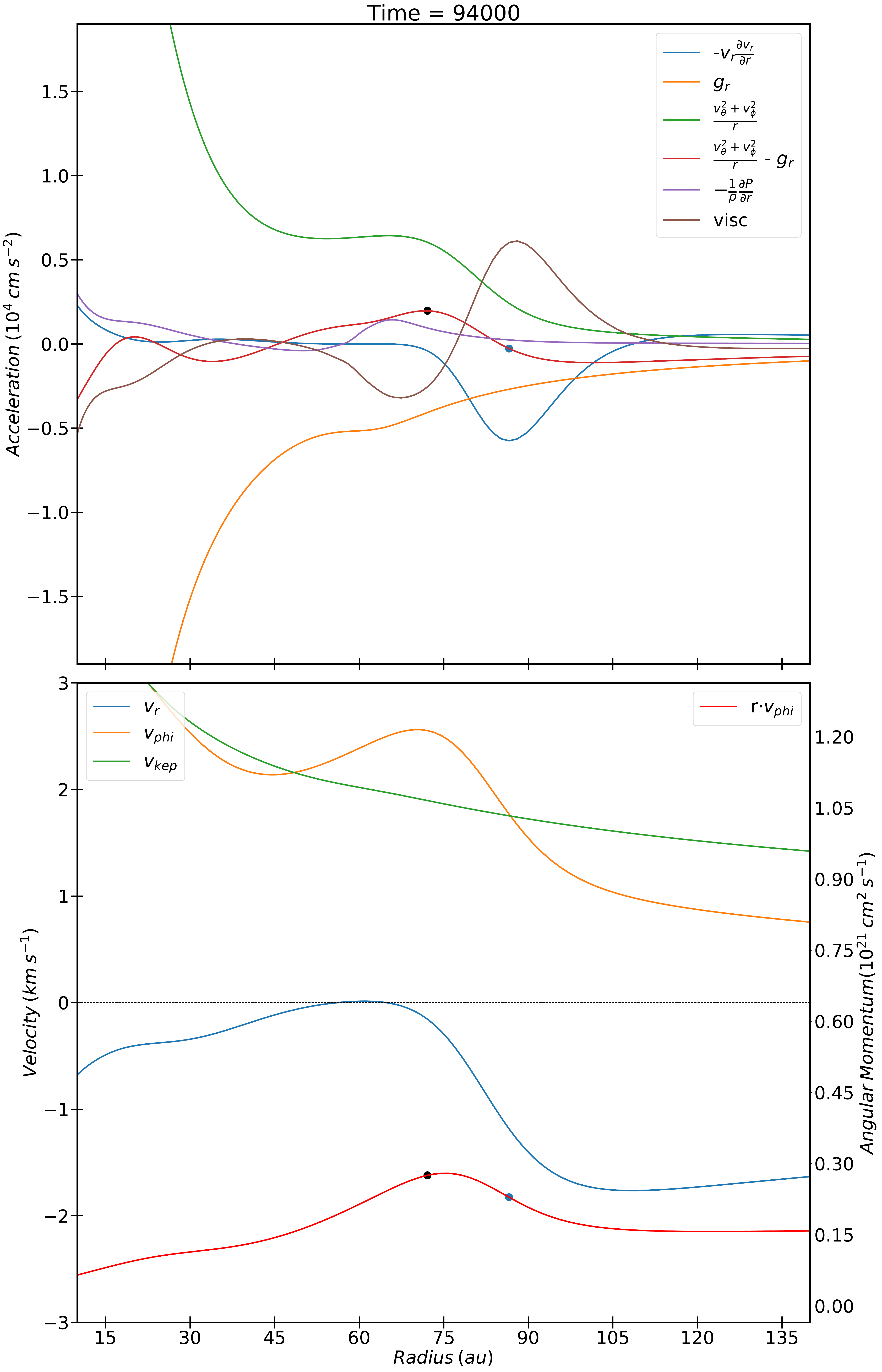}
        \label{fig:acag-second}
    \end{subfigure}
    \hfill
    
\caption{The top panels illustrate the gravitational acceleration (green), centrifugal acceleration (orange), pressure gradient (purple), difference between the centrifugal acceleration and the gravitational acceleration (red) in the radial direction, acceleration due to viscosity (brown) and y=0 (dashed black) at 91,500 (left panels) and 94,000 years (right panels). The lower panels display the rotational velocity (orange), infall velocity (blue), and Keplerian velocity (green) in $\mathrm{km\,s^{-1}}$. The specific angular momentum (red) is displayed containing dots which mark the radius with the maximum rotational velocity (black, inner dot) and the centrifugal radius, where the gravitational acceleration is balanced by the centrifugal acceleration (blue outer dot). On the upper panel, these dots are on the excess centrifugal force line. The specific angular momentum curve is in units of $10^{21}\,\mathrm{cm^2\,s^{-1}}$.
}
    
\label{fig:acag}
\end{figure*}

As discussed in \S~\ref{sec:intro}, at the heart of the concept of centrifugal barrier lies the question of how the rapid, supersonic infall motion is decelerated. If the infall speed reaches a maximum at the centrifugal radius $r_\mathrm{CR}$ where the gravitational acceleration is balanced by the centrifugal acceleration and the deceleration inside $r_\mathrm{CR}$ is dominated by the centrifugal force (rather than, say, pressure gradient), then the concept of centrifugal barrier would remain broadly valid. To address this question quantitatively, we will focus on two representative times, $t=91,500$ and $94,000$~years, when a well-defined rotationally supported disk has formed. In the lower panels of Fig.~\ref{fig:acag}, we plot as a function of radius (on a linear scale) the distributions of the radial velocity $v_r$ and rotational velocity $v_\phi$, the difference between the centrifugal acceleration and the gravitational acceleration, and the acceleration due to pressure gradient on the equatorial plane. 

The locations where the centrifugal acceleration balances the inwards gravitational acceleration are visible by the intersection of their respective difference, illustrated by the red dotted line in the upper panels of Fig.~\ref{fig:acag}, with the $x-$axis (the horizontal dashed black line). At time $t=94,000$~years (right panels), the first location of this intersection is $r \approx 7\times 10^{14}$~cm while the second occurs at $r \approx 1.3\times 10^{15}$~cm. The second location is where the infalling material reaches the Keplerian rotation for the first time, and can thus be identified as the centrifugal radius $r_\mathrm{CR}$. The deceleration region at this time is where the term $-v_r\ (\partial v_r/\partial r)$ (plotted as the blue dashed line in the upper panel) becomes negative, which ranges from $1.15\times 10^{15}$~cm to $1.6\times 10^{15}$~cm. Clearly, the radial deceleration started at a radius outside the Keplerian radius $r_\mathrm{CR}$, indicating that the deceleration is not initiated by the centrifugal force overtaking the gravity, unlike the classic picture of centrifugal barrier formation. 

In addition to the centrifugal force, there are two new forces absent in the classic (test-particle) picture that can decelerate the infalling material. The more obvious one is the gas pressure gradient, which is shown in the upper panels of Fig.~\ref{fig:acag} as a dot-dashed purple line. Although its value is somewhat less than the excess of the centrifugal acceleration over the gravitational acceleration (dotted red line) at $t=94,000$~years (see the upper-right panel), this is not the case at the earlier time $t=91,5000$~years (see the upper-left panel; see also the animated version of Fig.~\ref{fig:acag} in the supplementary material. At early times, the gas pressure gradient dominates the excess centrifugal force. As time progresses, the maximum excess centrifugal force over the gravity remains relatively unchanged as the peak location moves outwards with the growing disk, while the contribution of the gas pressure gradient decreases as time progresses. The presence of a significant pressure gradient makes it difficult to identify the super-Keplerian rotation found in the simulation as the classic centrifugal barrier. 



A second new ingredient that is absent from the classic picture is a means to redistribute or remove angular momentum, which is required to drive disk accretion. The most widely discussed mechanisms include magneto-rotational instability \citep[MRI,][]{1991ApJ...376..214B}, magnetically driven disk winds \citep{1982MNRAS.199..883B}, and gravitational torques for self-gravitating disks \citep{2016ARA&A..54..271K}, although the exact mechanism for driving the protostellar disk accretion remains unclear. For simplicity, in our exploratory (non-magnetic) axisymmetric simulation, we have adopted the standard $\beta-$prescription for the effective viscosity to drive the disk accretion (the viscosity is applied throughout the simulation domain). This viscosity introduces a force in the radial direction, which gives rise to an acceleration that helps to decelerate the infalling material in the envelope-to-disk transition zone (see the dot-dashed brown line in Fig.~\ref{fig:acag} and the Appendix on how this force is computed).  Indeed, at both times shown in the figure, the effective viscosity plays a more important role than the excess centrifugal force (over gravity) and pressure gradient in the initial deceleration of the infalling envelope outside the rotationally supported disk. The (outward) excess centrifugal force and pressure gradient become more important closer to the outer edge of the disk, where the bulk of the infall deceleration has completed and the force associated with the effective viscosity becomes negative (pointing inward). It appears that both the excess centrifugal force and pressure gradient play a role in slowing down the envelope infall, but in a complex way mediated by the viscosity. This complexity is evidence that the super-Keplerian rotation in our simulation is different from that envisioned in the classic picture of centrifugal barrier. 

Additional support for the above conclusion comes from a comparison of centrifugal radius $r_\mathrm{CR}$ and the radius of maximum super-Keplerian rotation $r_\mathrm{sK}$. At the representative time $t=94,000$~years shown in the right panels of Fig.~\ref{fig:acag}, we have $r_\mathrm{CR} = 1.253\times 10^{15}$~cm and $r_\mathrm{sK}=1.097\times 10^{15}$~cm, which yield a ratio $r_\mathrm{sK}/r_\mathrm{CR}$ of $0.8754$ that is substantially larger than the ratio of the classic centrifugal barrier to the centrifugal radius $r_\mathrm{CB}/r_\mathrm{CR}=0.5$. The same is true at the earlier time $t=91,500$~years shown in the left panels of Fig.~\ref{fig:acag}, where $r_\mathrm{sK}/r_\mathrm{CR}=0.8900$. Therefore, the maximum super-Keplerian rotation is reached closer to the centrifugal radius in our simulation compared to the classic (test-particle) picture, which casts a strong doubt on identifying the radius $r_\mathrm{sK}$ as the classic centrifugal barrier $r_\mathrm{CB}$. 

Another doubt on identifying $r_\mathrm{sK}$ as $r_\mathrm{CB}$ comes from the distribution of specific angular momentum on the equatorial plane. In the classic picture, the specific angular momentum of the test particle is conserved between the centrifugal radius $r_\mathrm{CR}$ and the centrifugal barrier $r_\mathrm{CB}$. However, this is not the case between $r_\mathrm{CR}$ and $r_\mathrm{sK}$ in our simulation. As illustrated by the green dotted line in the lower panels of Fig.~\ref{fig:acag}, the specific angular momentum is variable between $r_\mathrm{CR}$ and $r_\mathrm{sK}$, with the value at $r_\mathrm{sK}$ significantly higher than that at $r_\mathrm{CR}$. This would be surprising if the super-Keplerian region is part of the infalling envelope, which is expected to lose specific angular momentum as it collapses closer to the central object and spins up because the effective viscosity tends to remove angular momentum from the faster spinning material closer in and transfer it to the more slowly rotating material further out. This apparent dilemma disappears if the super-Keplerian is (the outer) part of the disk, where the angular momentum must be transported outward by the effective viscosity in order for the disk to accrete in the first place. In this case, the question becomes: why does the outer disk rotate at a super-Keplerian speed, unlike the rest of the disk?

The exact answer to the above question is unclear. It is likely related to the fact that the outer disk is in direct contact with the rapidly infalling envelope. One possibility is that the pressure gradient and the radial force associated with the effective viscosity are not strong enough to bring the rapid infalling envelope material to a complete stop in the radial direction when it enters the disk so that a super-Keplerian rotation is needed to give the material an extra outward push. This possibility was in fact anticipated by \citet{cassen1981formation}, who estimated the centrifugal force needed to balance the ram pressure generated by the infalling envelope (see their Appendix). The amount of extra centrifugal force (and associated super-Keplerian rotation) needed is controlled by complex interplay between angular momentum redistribution and radial force balance that is difficult to quantify. One thing we know for sure is that it depends on the magnitude of the viscosity, as discussed in \S~\ref{sec:viscosity}.
We will return to a discussion of the origin of the super-Keplerian rotation towards the end of \S~\ref{sec:viscosity}.



\subsection{Dependence of Super-Keplerian Rotation on Viscosity}
\label{sec:viscosity}

To explore the dependence of the super-Keplerian rotation on viscosity, we adopt an initial singular isothermal density profile following a $1/r^2$ power law. The simulation domain is the same as the reference model, but with an enclosed mass of $1.25~\mathrm{M_\odot}$ , an initial solid body rotation of $\Omega = 1.189 \times 10^{-13}~\mathrm{rad\,s^{-1}}$ corresponding to a ratio of rational and gravitational energies $\beta_\mathrm{rot} \approx 3.2\%$. The following range of viscosity parameters were explored: $\beta=0.001, 0.003, 0.01, 0.03, 0.1, 0.3, 1.0$. 

\begin{figure}
    \centering
    \includegraphics[width=\columnwidth]{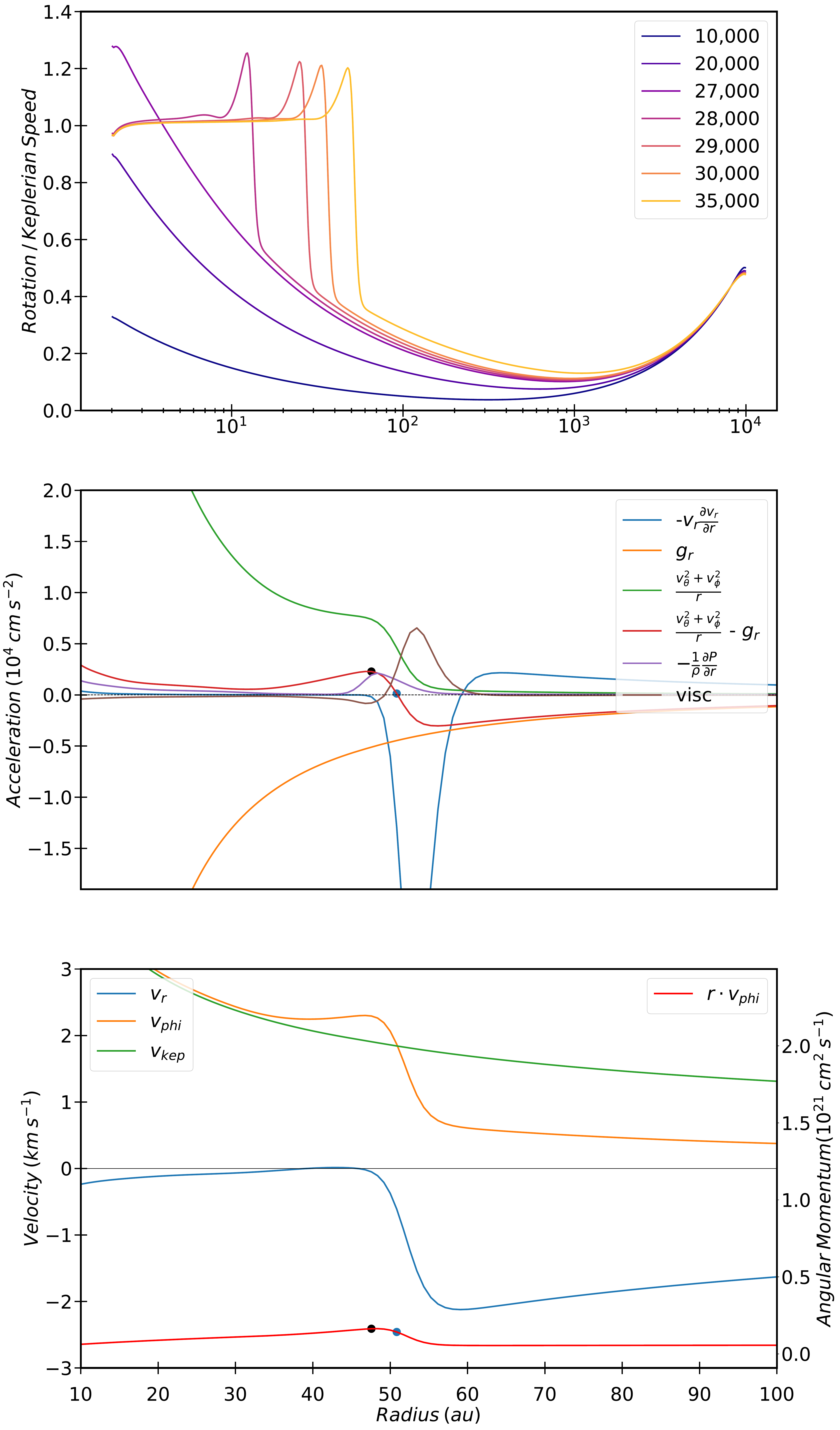}
    \caption{The top panel shows the evolution of super-Keplerian rotation for the singular isothermal sphere initial density profile with a viscosity parameter $\beta=0.03$ for selected time frames. The middle panel illustrates the gravitational acceleration (green), centrifugal acceleration (orange), pressure gradient (purple), difference between the centrifugal acceleration and the gravitational acceleration (red) in the radial direction, acceleration due to viscosity (brown) and y=0 (dashed black) at a time of 35,000 years. The lower panel displays the rotational velocity (orange), infall velocity (blue), and Keplerian velocity (red) in $\mathrm{km\,s^{-1}}$}
    \label{fig:three_panel}
\end{figure}

We start the discussion in the middle of the parameter range, with the $\beta=0.03$ viscosity model. This run is broadly similar to the previously described reference run, as illustrated in Fig.~\ref{fig:three_panel}, where we plot the evolution of the ratio of the rotation speed to the local Keplerian speed on the equatorial plane at different times (top panel), accelerations due to various forces (middle panel, as in the upper panels of Fig.~\ref{fig:acag}) and the same 4 quantities as in the lower panels of Fig.~\ref{fig:acag} (bottom panel) at a time when the disk radius is about 100~au (or $1.5\times 10^{15}$~cm).

The top panel shows the similarities between this model with an initial singular isothermal sphere density distribution and our reference run in terms of the evolution of super-Keplerian rotation. Early on in the model gas spins up faster at smaller radii, until around 27,000 years where a maximum rotational velocity is reached, corresponding to a ratio value of 1.28 followed immediately by disk formation. The maximum rotational speed towards the outer edge of the disk starts with a ratio value of 1.26, but decreases as the disk propogates outwards to a final value of 1.15 in the last time frame shown.  It has an excess centrifugal acceleration (over gravity) that is not dominant in decelerating the infalling envelope and a specific angular momentum that is significantly higher than that in the envelope, both of which point to an origin other than the centrifugal barrier. Our conclusions from the reference case are thus strengthened. 

\begin{figure}
    \centering
    \includegraphics[width=\columnwidth]{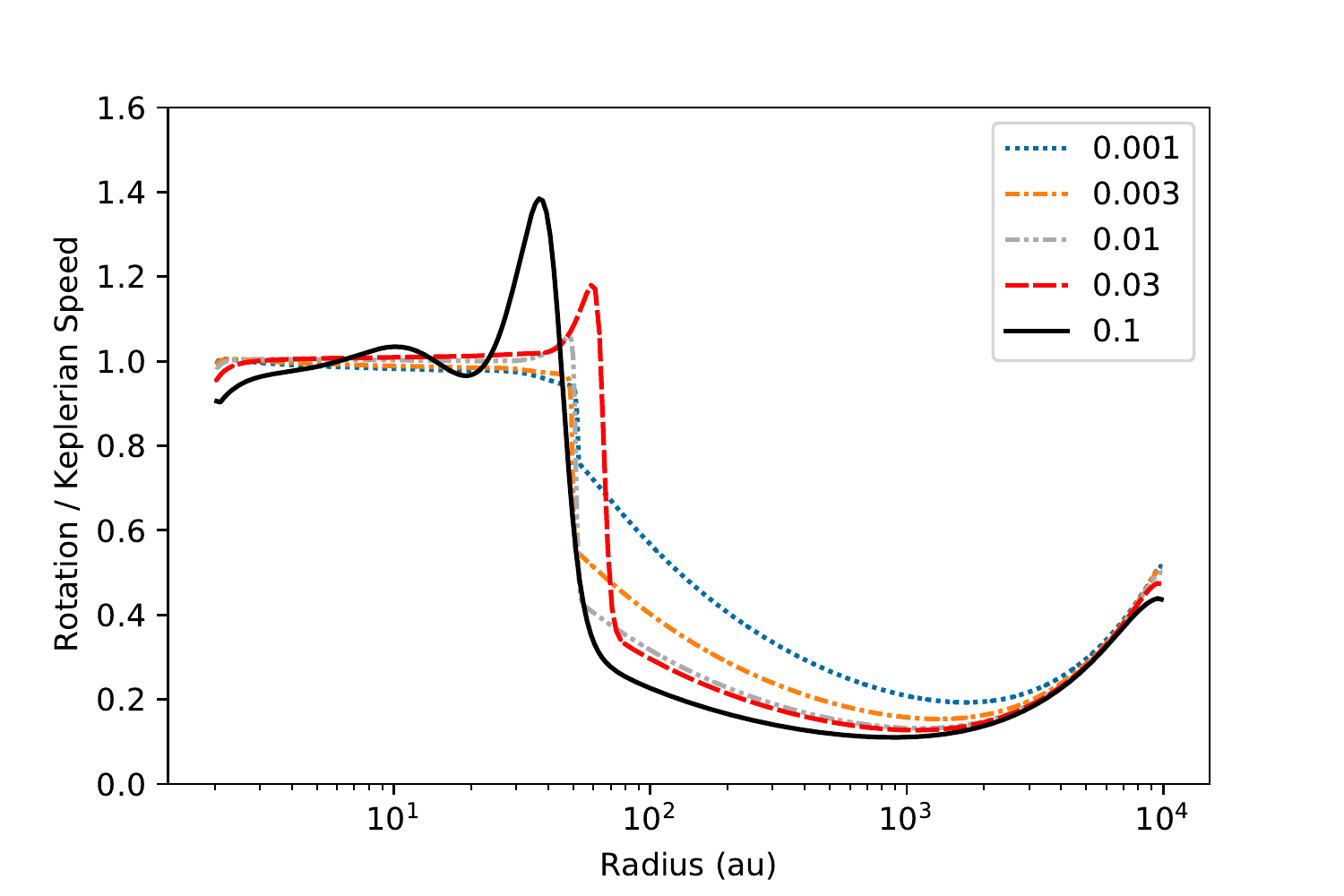}
    \caption{Ratios of rotational speed to the local Keplerian speed for selected viscosity values, showing that the super-Keplerian rotation becomes more prominent with increasing viscosity.}
    \label{fig:all_sis_beta}
\end{figure}

To illustrate the effects of viscosity on super-Keplerian rotation, we plot in Fig.~\ref{fig:all_sis_beta} the ratios of rotation speed to the local Keplerian speed around the time when the disk radius is about 100~au for different values of the viscosity parameter $\beta$. A clear trend can be seen within Fig.~\ref{fig:all_sis_beta} - increased viscosity tends to lead to a faster rotation speed compared to the local Keplerian speed near the disk outer edge up to the $\beta$ = 0.1 model. Lower viscosity models are able to form disks with rotation speeds closer to the Keplerian speed. A visible super-Keplerian region starts to appear at $\beta$ = 0.01. It grows in magnitude as the $\beta$ viscosity increases, forming a prominent super-Keplerian rotating region with the $\beta$ = 0.1 model. We have explored models with even higher viscosities, such as $\beta = 0.3, 1.0$, but they do not form disks, likely because angular momentum is redistributed too quickly in the rotating collapsing envelope to allow the disk to form. With these sets of models, we once again find that the location of the peak super-Keplerian rotation moves outwards while the material at this location  travels inwards, confirming the wave nature of the super-Keplerian region discussed in the reference model. 

The fact that the super-Keplerian rotation becomes more prominent with increasing viscosity suggests the following thought experiment that may shed light on its origin. Imagine a viscous disk in mechanical equilibrium that is contained within a fixed radius (say by the infalling envelope). If the viscosity is zero, the disk would remain in equilibrium without the need for adjustment. However, the presence of a viscosity would allow the faster rotating inner part of the disk to lose angular momentum (and accrete) and the more slowly rotating outer part of the disk to gain angular momentum, which would normally lead to expansion in the absence of any external confinement. In the limit that the disk is confined to a fixed radius, angular momentum would accumulate near the outer edge of the disk, eventually leading to a solid-body rotation (with $v_\mathrm{K}\propto r$ rather than $r^{-1/2}$) that minimizes the frictional force between adjacent disk annuli and thus the rate of angular momentum redistribution. The time scale to reach such a (super-Keplerian, asymptotic) solid-body rotation depends on the value of viscosity: it decreases as viscosity increases. This thought experiment leads us to the conjecture that the super-Keplerian rotation is caused by the outward angular momentum transport in the disk while the disk expansion is constrained by the infalling envelope. Specifically, it results from the disk not being able to expand fast enough to accommodate the angular momentum transported to the outer disk. 
 
\section{Conclusion and Discussion}
\label{sec:discussion}
 
 
 Motivated by recent ALMA kinematic observations of the embedded disk L1527 and their interpretation as evidence for the classic centrifugal barrier \citep{2014Natur.507...78S,sakai2017vertical}, we have carried hydrodynamic simulations of protostellar disk formation including a minimum set of ingredients: rotation, self-gravity and an effective viscosity to ensure disk accretion. The focus was on super-Keplerian rotation, which is a key characteristic of the centrifugal barrier. We indeed find a super-Keplerian region surrounding the Keplerian part of the disk (see Figs.~\ref{fig:vphi_ratio}, \ref{fig:three_panel} and \ref{fig:all_sis_beta}), with a maximum ratio of the rotation speed to the local Keplerian speed that can approach $\sqrt{2}$ (see the black curve in Fig.~\ref{fig:all_sis_beta}), the value predicted for the classic centrifugal barrier. However, unlike the classic picture, the infalling envelope is decelerated by a combination of the pressure gradient, centrifugal force, and the radial component of the force due to the effective viscosity (see the upper panels of Fig.~\ref{fig:acag}) rather than the centrifugal force alone. The maximum super-Keplerian rotation $r_\mathrm{sK}$ is located significantly closer to the Keplerian radius $r_\mathrm{CR}$ (where the rotating infall envelope material first reaches the local Keplerian speed) than the centrifugal radius $r_\mathrm{CB}$ (see the lower panels of Fig.~\ref{fig:acag}). Furthermore, the specific angular momentum at the location of maximum super-Keplerian rotation ($r_\mathrm{sK}$) is substantially higher than at the centrifugal radius $r_\mathrm{CR}$, indicating that the super-Keplerian region is a result of the (outward) angular momentum transport inside the disk (subject to the constraint imposed by the infalling envelope on the disk expansion), rather than simply a consequence of the envelope material spinning up to conserve angular momentum as it falls closer in. This interpretation is supported by simulations with different values of viscosity, where we find that the super-Keplerian rotation becomes more prominent with increasing viscosity. We conclude that super-Keplerian rotation can exist in protostellar disk formation, but it differs substantially from the classic picture of centrifugal barrier.
 
 We note that an enhanced rotation (and surface density) was also found by \cite{Saigo1998} near the outer edge of the rotationally supported structure in their self-similar solutions of the collapse of rotating, thin (sheet-like), isothermal clouds (see their Fig. 4 and 6). However, it is unclear whether the enhanced rotation near the disk outer edge is super-Keplerian or not, because the Keplerian speed could also be enhanced due to the enhanced surface density. Furthermore, their solutions did not include any angular momentum removal or redistribution mechanism, which yields a zero mass for the central stellar object, which, in turn, complicates a direct comparison with our simulations.
 Nevertheless, this idealized work highlights an important conceptual point: the location of the accretion shock that separates the rotationally supported (disk) structure and the infalling envelope (and the flow properties near the shock) is determined globally. Specifically, in a similarity solution, it is determined by matching at the (initially unknown) shock location an inner (disk) solution to the governing ordinary differential equations obtained by outward shooting from near the origin and an outer (envelope) solution obtained from inward shooting from the infinity. This global nature of the solution is in line with our conjecture that viscosity can change the disk properties near the outer edge (including super-Keplerian rotation) by controlling the inward mass accretion and outward angular momentum transport in the inner (disk) solution. This global nature of the problem, involving a complex interplay between the disk dynamics (particularly accretion and spreading) and the need for mechanical balance in the envelope-disk transition zone, makes it difficult to interpret the numerical results with certainty.

 Obvious future refinements of our exploratory work include the treatment of magnetic fields (and the associated non-ideal MHD effects) and extension to three dimensions (3D). Magnetic fields will enable the disk to accrete through the physical processes of magneto-rotational instability and/or disk-wind without a prescribed effective viscosity and 3D simulations are needed to capture the angular momentum transport through spiral density waves. Whether the super-Keplerian rotation found in our simulation persists when these and other physical effects are included remains to be determined. There is some anecdotal evidence that this may be true in at least some cases. For example, in one of the models presented in \citet{2018MNRAS.478.2723Z} that includes a relatively strong magnetic field (with a dimensionless mass-to-flux ratio of 2.4) and ambipolar diffusion in the absence of small grains, the outer part of the disk is rotating at a speed well above the Keplerian speed based on the central stellar mass (see their Fig.~15, left panel), although it is unclear whether the region remains super-Keplerian when the disk mass is included in the computation of the Keplerian speed. Additional work is needed to determine how common super-Keplerian rotation exists in well-resolved simulations that include more detailed physics (e.g., \citealt{Xue2021}).
 There are also examples where the disks formed in magnetized cores are clearly Keplerian or slightly sub-Keplerian \citep[see, e.g., Fig.~5 of][bottom panels]{2020ApJ...898..118H}. Additional work is needed to determine how common super-Keplerian rotation exists in well-resolved simulations that include more detailed physics. If our interpretation is correct, the  super-Keplerian rotation is expected to occur most likely in fastest-accreting youngest (Class 0) disks where angular momentum is transported outward most quickly outward and, at the same time, the disk expansion is most constrained by the infall of a massive envelope.  
 
 Ultimately, whether super-Keplerian rotation exists in a disk should be decided based on observations. This will not be an easy task, especially for deeply embedded protostellar disks that are still in active formation because it is difficult to determine the stellar mass (and the disk mass) needed for evaluating the Keplerian speed independent of the rotation speed (from line observations). It is easier to determine the mass of the optically visible central star of the more evolved protoplanetary disk, but such a disk is less likely to have super-Keplerian rotation if such a feature comes from the interplay between the outward angular momentum transport and the constraint of disk expansion by the infalling envelope. 
 
 One implication of the potential super-Keplerian rotation is that it would complicate the mass determination of protostars based on the disk rotation measurement. In particular, the rotation data in regions that rotate up to $\sqrt{2}$ times faster than the local Keplerian speed could lead to an over-estimate of the stellar mass by up to a factor of 2. Another implication is on the dynamics of dust grains, which tend to orbit at the local Keplerian speed and are thus expected to experience a tail-wind in the super-Keplerian region that may lead to an outward migration rather than the commonly expected inward radial drift. The difference may lead to modification to dust concentration and growth that deserves future investigations.

\section*{Data Availability}
The data underlying this article will be shared on reasonable request to the corresponding author.

\section*{Acknowledgements}

We thank N. Sakai and the referee for helpful comments. DCJ and KHL acknowledge support from NSF AST-1716259. ZYL is supported in part by NASA 80NSSC20K0533 and NSF AST-1910106. YT acknowledges support from VICO Interdisciplinary Fellowship from the University of Virginia.




\bibliographystyle{mnras}
\bibliography{biblio} 




\appendix

\section{Radial Acceleration From the Viscosity Term in Momentum Equation} 

 Here we provide the formulae to compute the radial acceleration on the equatorial plane from the viscosity term in the momentum equation \ref{eq:momentum},  $\nabla\cdot \vec{\Pi}$. From equation~(7) from \citet{2020ApJS..249....4S}, we have 
 $$
 \vec{\Pi}_{ij} = {\rho}\nu_\beta \left\{{\partial v_i\over\partial x_j} + {\partial v_j\over\partial x_i} - \frac{2}{3} \delta_{ij} \nabla \cdot \vec{v} \right\}
 \equiv   \tau_{ij} - P_{\beta} \delta_{ij}
$$
where 
$$
P_\beta\equiv \frac{2}{3} {\rho}\nu_\beta \nabla \cdot \vec{v} 
$$ 
is an effective pressure term associated with the viscosity tensor and $\tau_{ij}$ denotes the viscosity tensor excluding the effective pressure term. 

In a spherical polar coordinate system under the assumption of axisymmetry, the radial component of the momentum equation~(\ref{eq:momentum}) becomes:
 
 $$
 {\rho} \left\{ {\partial v_r\over\partial t}+v_r{\partial v_r\over\partial r}+{v_\theta\over r}{\partial v_r\over \partial \theta}-{v_\phi^2 + v_\theta^2\over r} \right\} = -{\partial P\over\partial r} -{\partial P_\beta\over\partial r} +
 $$
 $$
 \left\{ {1\over r^2}{\partial (r^2\tau_{rr})\over\partial r} + {1\over r \sin\theta}{\partial (\tau_{\theta r}\sin\theta)\over \partial \theta} -{\tau_{\theta\theta}+\tau_{\phi\phi}\over r }  \right\} + {\rho}g_r
 $$
 
 where
 $$
 P_\beta={2\over 3}({\rho\nu_\beta}) \left[ {1\over r^2}{\partial\over\partial r}(r^2 v_r) + \frac{1}{v\sin\theta}\frac{\partial v_\theta\sin\theta}{\partial\theta} \right]
 $$
 and $\nu_\beta$ is the kinematic viscosity that depends on $\beta$, $\Omega$ and $R$ through equation~(\ref{eq:viscosity}). 
 
 The relevant elements of the tensor $\tau_{i,j}$ are
 $$
 \tau_{rr}=[\rho \nu_\beta] \left(2{\partial v_r\over \partial r}\right)
 $$
 
 $$
 \tau_{\theta\theta}=[\rho\nu_\beta] 2 \Big(\frac{1}{r} \left({\partial v_\theta\over \partial \theta}\right) + \frac{v_r}{r}\Big)
 $$
 
 $$
  \tau_{\phi\phi} = [\rho\nu_\beta] 2 \Big(\frac{v_{\theta} cot(\theta)}{r} + \frac{v_{r}}{r}\Big)
 $$

 $$
  \tau_{\theta r} = [\rho\nu_\beta] \Big(r \frac{\partial}{\partial r} \Big(\frac{v_{\theta}}{r}\Big) + \frac{1}{r} \frac{\partial v_r}{\partial \theta} \Big)
 $$


\bsp    
\label{lastpage}
\end{document}